\documentclass[aps,prb,twocolumn,showpacs,superscriptaddress,floatfix]{revtex4}
\usepackage{graphicx}
\usepackage{float}
\usepackage{xcolor}
\usepackage{amsmath}
\usepackage{braket}
\usepackage{caption}
\usepackage{subcaption}
\newcommand{\up}{\uparrow}
\newcommand{\dn}{\downarrow}
\usepackage[justification=Justified,format=plain]{caption}
\begin{document}
\title{Non-perturbative indirect exchange in spin-valley coupled 2D crystals}

\author{M. R. Losada}
\affiliation{Departamento de F\'isica  At\'omica, Molecular y  Nuclear, Universidad de Granada, 18071 Granada, Spain} 

\author{A. T. Costa}
\affiliation{
 International Iberian Nanotechnology Laboratory, 
Av. Mestre Jos\'e \ Veiga,\ 4715-330 \ Braga, \ Portugal 
}
\author{B. Biel}
\affiliation{Departamento de F\'isica  At\'omica, Molecular y Nuclear, Universidad de Granada, 18071 Granada, Spain}
\affiliation{Instituto Carlos I de F\'isica Te\'orica y Computacional, Universidad  de  Granada, 18071 Granada, Spain
}

\author{J. Fern\'andez-Rossier\footnote{On leave from Departamento de F\'isica Aplicada, Universidad de Alicante, Spain
}}
\affiliation{
 International Iberian Nanotechnology Laboratory, 
Av. Mestre Jos\'e \ Veiga,\ 4715-330 \ Braga, \ Portugal 
}

\date{\today}

\begin{abstract}
We study indirect exchange interactions between localized spins of magnetic impurities in spin-valley coupled systems described with the Kane-Mele model. Our model captures the main ingredients of the energy bands of 1H transition metal dichalcogenides (TMDs) monolayers, such as 1H-MoS$_2$ and 1H-NbSe$_2$. To obtain the effective interactions, we use the exact diagonalization of the Hamiltonian, avoiding momentum cut-offs. We start by comparing the standard perturbation expansion in terms of the Kondo exchange with the exact calculation of the interaction, treating the local spins classically. We find that perturbation theory works well even beyond the regime where the relevant figure of merit, the ratio between the exchange $J$ and the hopping $t$, is small. We verify that the effective indirect exchange Hamiltonian derived from perturbation theory also works in the non-perturbative regime. Additionally, we analyze the interplay between the symmetry of the different terms of the interaction (Heisenberg, Ising, and Dzyaloshinskii–Moriya (DM)), the Fermi-surface topology, and the crystallographic direction in which the impurities are placed. We show that the indirect exchange along the armchair direction is actually Heisenberg-like, due to the reflection symmetry of the crystal structure around this direction. Finally, we explore the exploitation of indirect exchange, combined with atomic manipulation, to engineer the Majumdar-Ghosh Model. Our results show that TMDs provide an extremely versatile platform to engineer indirect exchange interactions.
\end{abstract} 

\maketitle

\section{INTRODUCTION}
The effective coupling that arises when two otherwise-decoupled local spins interact with the same electron gas is known as indirect exchange interaction. Indirect interaction was studied for the first time by Ruderman and Kittel\cite{PhysRev.96.99} for the case of nuclei. Then, Kasuya and Yosida \cite{kasuya1956theory,yosida1957magnetic} realized that the same kind of physics applies to magnetic impurities, and this is the application that has turned out to be more relevant over the years. The Ruderman-Kittel-Kasuya-Yosida (RKKY) interaction is mediated by the conduction electrons, and, as such, it is determined by the spin susceptibility of the host material. Indirect exchange interaction was first considered in the case of metals, where the RKKY interaction decays and oscillates with the distance between the impurities. Its strength depends on the dimension of the host material and on the actual magnetic species, whereas the oscillation period is determined by the Fermi surface. Thus, peculiar Fermi surfaces are expected to give rise to non-conventional indirect exchange interactions \cite{PhysRev.149.519}. 

This brings us to consider one of the most intriguing materials of the past few years: graphene\cite{novoselov2004electric,RevModPhys.81.109}. RKKY interaction in graphene has been thoroughly studied\cite{PhysRevB.74.224438,PhysRevLett.99.116802,PhysRevB.76.184430,black10,PhysRevB.84.125416,sherafati11b,kogan2012while} and, not surprisingly, it is predicted to have properties significantly different from the typical behavior in metals. In the first place, the decay of the interaction with the separation between impurities is cubic ($R^{-3}$), different from the $R^{-2}$ found in 2D metals\cite{PhysRevB.11.2025}. On the other hand, due to the bipartite nature of the graphene honeycomb lattice, the preferred alignment of the spin of the impurities depends on whether both impurities are placed on the same sublattice or not\cite{PhysRevLett.99.116802}. 

A related class of 2D materials that can have peculiar Fermi surfaces are transition metal dichalcogenides, such as MoS$_2$ and NbSe$_2$. In these materials, the combination of strong spin-orbit coupling and the lack of inversion symmetry gives rise to a spin splitting in the energy bands, with a maximum of the splitting placed at the Dirac points, which leads to a spin-valley coupled band structure\cite{xiao2012coupled,kosmider13}. The peculiar Fermi surface of these materials affects directly the indirect exchange interaction\cite{hatami2014spin,parhizgar2013indirect,mastrogiuseppe2014rkky}. In contrast to the case of graphene, indirect exchange in TMDs is no longer of Heisenberg type but consists of three different coupling terms: Heisenberg, Dzyaloshinsky-Moriya, and Ising \cite{parhizgar2013indirect}, on account of the spin-orbit coupling (SOC) and lack of inversion symmetry.

In the context of TMDs, indirect exchange can be relevant in at least three different scenarios. First, magnetic dopants that substitute either the transition metal or the chalcogen atom \cite{ramasubramaniam13,menard15,catarina20,zhang2020,pham2020,saha21}, 
forming a diluted magnetic semiconductor. Second, compounds such as Cr\textsubscript{1/3}NbS\textsubscript{2}, where Cr atoms are intercalated in between the TMDs planes\cite{moriya82} and helical magnetic order has been reported\cite{miyadai1983magnetic,tsuruta2016phase,togawa2012chiral,lu2020canted}. And third, in those TMDs-based van der Waals heterostructures where a conducting layer is adjacent to an insulating layer that hosts spins, such as 1T/1H TaS\textsubscript{2}\cite{vavno21,ayani22}. In this paper, we will address the first scenario. 

Previous work on RKKY interactions in TMDs\cite{hatami2014spin,parhizgar2013indirect,mastrogiuseppe2014rkky} is based on the $k\cdot p$ model\cite{xiao2012coupled}, and treats the exchange between itinerant electrons and the local spins perturbatively. This $k\cdot p$ approach requires the introduction of a momentum cut-off that translates into low resolution in the indirect exchange interaction in real space and gives rise to some discrepancies between the different studies \cite{kogan2012while}. Additionally, in some cases\cite{hatami2014spin}, intervalley scattering is neglected, resulting in a vanishing Dzyaloshinskii–Moriya. 
  
In this work, to overcome these drawbacks, we compute the effective coupling through the exact diagonalization of a lattice Hamiltonian in a large simulation cell with periodic boundary conditions. Therefore, we do not need to rely on the $k\cdot p$ approach and our calculation is not perturbative. This approach also allows us to validate the results obtained from second-order perturbation theory. Furthermore, we show that the effective Hamiltonian derived from perturbation theory describes the indirect exchange in the non-perturbative range. 

In order to provide meaningful results for realistic systems, the parameters of our Hamiltonian have been set to capture the characteristics of 1H TMDs. We have thus computed the RKKY interaction along the high-symmetry directions of those materials: armchair and zigzag. We focus on the hole-doped case, which is expected to be relevant because of the larger spin-splitting of the valence bands in TMDs. By performing the Fourier transform of the effective coupling, we can identify the different scattering processes that contribute to the interaction. Finally, we show how to take advantage of the richness of the interplay of physical phenomena in this system to construct the Majumdar-Ghosh state\cite{majumdar1970antiferromagnetic}.

\section{METHOD AND THEORY}
\subsection{Kane-Mele model}
\begin{figure}[t]
    \begin{subfigure}[r]{0.45\columnwidth}
    \includegraphics[width=4cm]{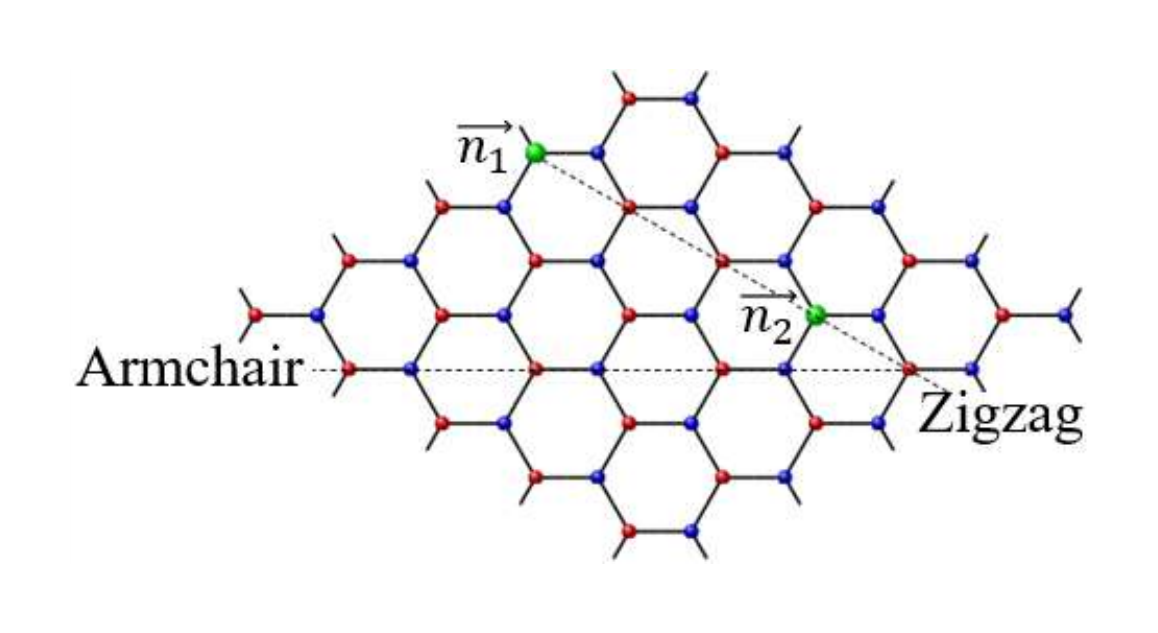} 
    \text{\quad \qquad \quad       (a)}
    \label{fig:lattice}
    \end{subfigure}
    \quad 
    \begin{subfigure}[l]{0.45\columnwidth}
    \includegraphics[width=4.2cm]{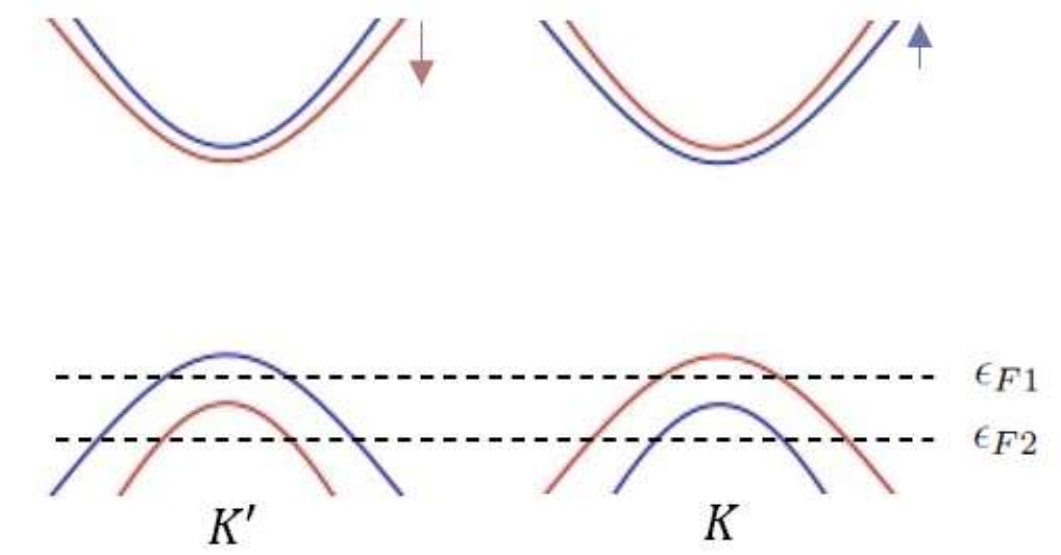}
    \text{\quad (b)}
    \label{fig:band}
    \end{subfigure}
    \begin{subfigure}{0.3\textwidth}
    \includegraphics[width=5cm]{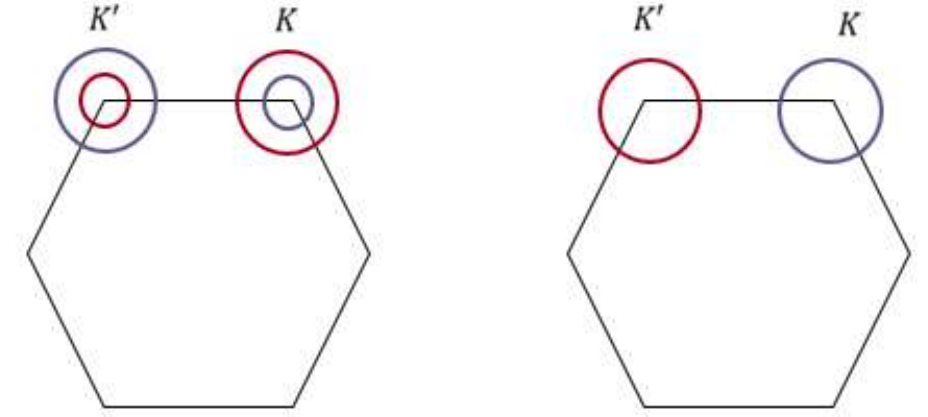}
    \text{\quad  (c)}
    \label{fig:top}
\end{subfigure}

\caption{(a) Honeycomb lattice, red circles indicate sublattice A while blue circles correspond to sublattice B. Green circles represent the magnetic impurities positioned on sublattice A. When considering TMDs, the red sublattice indicates the transition metal elements, and the blue circles mark the position of the dichalcogenides. (b) Schematic low-energy band structure at $K$ and $K'$ valleys. The blue curves correspond to spin-up eigenstates and the red ones to spin-down. (c) First Brillouin zone. The red and blue circles correspond to the different Fermi surfaces. The right (left) picture shows the Fermi surface when the Fermi level is set to $\epsilon_{F1}$ ($\epsilon_{F2})$.}
\label{fig:system_bands}
\end{figure}

As our formal basis for the description of spin-valley coupled systems, we adopt the Kane-Mele model\cite{kane2005quantum}, which describes spinful fermions in a graphene-like lattice model [Fig.\,\ref{fig:system_bands}(a)]. In this model, the Hamiltonian reads as: 
\begin{equation}
    {\cal H}_0 = t \sum_{\langle ij \rangle,\sigma} c^{\dagger}_{i\sigma}c_{j\sigma} 
    + \sum_{i,\sigma} \epsilon_i c^{\dagger}_{i\sigma}c_{i\sigma}+
    i t_{KM}\sum_{\langle \langle ij  \rangle \rangle,\sigma} \nu_{ij}c^{\dagger}_{i\sigma}s^{z} c_{j\sigma},
    \label{H0}
\end{equation}
where $c_{i\sigma}^{\dagger}$ and $c_{i\sigma}$ are the creation and annihilation fermionic operators at site $i$ with spin $\sigma = \{\uparrow,\downarrow\}$. The first term describes the nearest-neighbor hopping with amplitude $t$. For the onsite term, we take $\epsilon_i=\pm\frac{\Delta}{2}$, with positive (negative) sign for the $A$ ($B$) sublattice. The third term describes the spin-orbit coupling, which involves a spin-dependent second neighbor complex hopping with amplitude $t_{KM}$, proposed by Kane and Mele \cite{kane2005quantum}. Here, $\nu_{ij} = \pm 1$ depends on the orientation between the bonds that connect the second neighbor sites. 

When $\epsilon_i = t_{KM} = 0 $, the Hamiltonian of Eq.\,(\ref{H0}) becomes the conventional one-orbital first-neighbor tight-binding model for graphene\cite{RevModPhys.81.109}, which gives rise to the zero-gap semiconductor band structure with Dirac cones. The onsite energy term breaks the inversion symmetry, and, as a result, a trivial gap opens at the corners $K$ and $K'$ of the first Brillouin zone. Because of that, the valence and conduction band states in the neighborhood of the $K$ and $K'$ points are mostly localized in the $A$ and $B$ sublattices, respectively. Now, if we add the SOC term, the otherwise spin-degenerate bands split into two spin-polarized bands\cite{costa22}. Due to time-reversal symmetry, Krammer’s doublets have opposite spin and momentum. As a result, the spin-splitting of the bands is opposite in each valley. This kind of band structure [Fig.\,\ref{fig:system_bands}(b)] leads to the so-called spin-valley coupled systems\cite{xiao2012coupled}. In the case of TMDs, the strong SOC comes from the orbital of the transition metal. In the tight-binding model, we describe it by considering different $t_{KM}$ for each sublattice. As a result, the spin-split of the valence band is much larger than the split of the conduction band.

\subsection{Exchange interaction between local spins and itinerant electrons}
Now we consider the interaction between the magnetic impurities and the conduction electrons of the host material. This kind of interaction is usually described through a contact Hamiltonian of the form: 
\begin{equation}
    {\cal V}_{int} = \sum_i J_i \vec{n}_i \cdot \vec{s}(\vec{r}_i),
    \label{Kondo1}
\end{equation}
where $J_i$ is the coupling constant between conduction electrons and impurities, $\vec{n}_i$ is the magnetic moments of the impurities, considered as classical vectors, and $\vec{s}(\vec{r}_i)$ is the spin density operator at site $\vec{r}_i$. We shall refer to Eq.\,(\ref{Kondo1}) as {\em Kondo} interaction, although we must note that we are treating local spins classically and therefore we can not describe the Kondo effect. 

\subsection{Indirect exchange interaction}
These local Kondo interactions lead to non-local effective or indirect coupling between distant local spins. The most accurate effective spin-spin interaction is obtained from directly calculating the energy difference between the parallel and anti-parallel configurations. Because of the anisotropy, this procedure is implemented along the three Cartesian indexes. As a result, the effective interaction depends on the orientation of the classical magnetization of the impurities:
\begin{equation}
   {J}_\mathrm{{eff}}(\vec{n}_1,\vec{n}_2) = \frac{E(\vec{n}_1,\vec{n}_2)
   -E(\vec{n}_1,-\vec{n}_2)}{2},
   \label{jeff}
\end{equation}
where $E(\vec{n}_1,\vec{n}_2)$ is the total energy of the systems computed by numerical diagonalization of ${\cal H}_0+{\cal V}_{int}$ in a large simulation cell with two magnetic impurities. We verified that our simulation cells are large enough so that our results do not depend on their size. 

\subsection{Perturbative Indirect-exchange}

A good starting point for the perturbative approach is to define the non-local spin susceptibility of the electrons. It averages spin density in position $\vec{r}_2$ due to the presence of one impurity placed at $\vec{r}_1$: 

\begin{equation}
\langle S_{\alpha}(\vec{r}_2) \rangle = J\sum_{\beta} \chi_{\alpha\beta}(\vec{r}_1-\vec{r}_2) n_\beta(\vec{r}_1),
\label{suscdef}
\end{equation}
where $\alpha,\beta = {x,y,z}$. Using linear response theory, the non-local spin susceptibility can be expressed (see Appendix) in terms of the eigenstates and eigenvalues of the single-impurity Hamiltonian ${\cal H}_0$. Here, we take a different approach and we determine it numerically, by directly computing the change of the spin density solving:   
\begin{equation}
    {\cal H}_1|\psi_n\rangle=\epsilon_n|\psi_n\rangle,
\end{equation}
where ${\cal H}_1={\cal H}_0+{\cal V}_{\rm int}$ is the Hamiltonian
for the system with one impurity at $\vec{r}_1$. Then we compute:
\begin{equation}
     \langle S_{\alpha}(\vec{r}_2) \rangle=\sum_n f_n 
      \langle\psi_n| S_{\alpha}(\vec{r}_2)|\psi_n \rangle,
      \label{eqn:spin_density}
\end{equation}
where $f_n=f(\epsilon_n)$ is the Fermi-Dirac function at $T=0$. Since we use a simulation cell with periodic boundary conditions, the resulting susceptibility depends on the relative position $\vec{r}_2-\vec{r}_1$ and we can take the impurity at the origin without loss of generality. We can then verify, numerically, that the expectation value of the spin density scales linearly with $J$, and pull out the susceptibility matrix from Eq.\,(\ref{suscdef}). 

We note that, for a given pair of sites, $\vec{r}_1,\vec{r}_2$, the non-local susceptibility is a three-by-three matrix, on account of the spin components $\alpha,\beta$. As we are dealing with a system that presents broken inversion symmetry and strong SOC, our susceptibility matrix is no longer diagonal but has the following structure:

\begin{equation}
\begin{pmatrix}
 \chi_{xx} & \chi_{xy} & 0\\
-\chi_{xy} &  \chi_{xx} & 0 \\
0 & 0 & \chi_{zz}.
\end{pmatrix}
\end{equation}

Once the non-local susceptibility is obtained, the indirect exchange Hamiltonian can be obtained 
assuming that the second classical impurity at site $\vec{r}_2$ interacts with the spin density induced by the first impurity placed at $\vec{r}_1$ via Eq.\,(\ref{Kondo1}). This leads to: 
\begin{equation}
\label{eqn:hamil_rkky}
    H_{RKKY} = J^2 \sum_{\alpha,\beta =x,y,z} n^1_{\alpha}\chi_{\alpha \beta}(\vec{r}_1-\vec{r}_2) n^2_{\beta}.
\end{equation}
This equation can also be obtained from treating Eq.\,(\ref{Kondo1}) to second-order in perturbation theory.
Equation (\ref{eqn:hamil_rkky}) naturally leads to the definition of three important energy scales:
\begin{eqnarray}
J_{xx} = J^2 \chi_{xx} \nonumber\\
J_{xy} = J^2 \chi_{xy}\nonumber\\
J_{zz} = J^2 \chi_{zz},  
\end{eqnarray}
and to an effective Hamiltonian
with  three types of coupling, Heisenberg, Ising, and DM: 
\begin{equation}
H= J_H \vec{n}_1\cdot\vec{n}_2
+ J_I n_1^z n_2^z +
J_{DM} (\vec{n}_1\times\vec{n}_2)\cdot\hat{z},
\label{eff_rkky}
\end{equation}
where $J_H=J_{xx}$ and $J_I=(J_{zz}-J_{xx})$, $J_{DM}=J_{xy}$. Thus, the RKKY interaction in spin-valley coupled system has two in-plane components, $J_{xx}$ and $J_{xy}$, and one out-plane component $J_{zz}$. While $J_{xx}$ and $J_{zz}$ favor a collinear alignment of the spins, the DM coupling favors a perpendicular alignment, and their competition leads to spin canting.  

\section{Validity of perturbative theory}

In this section, we use the non-perturbative approach to test the validity range of perturbation theory. We also verify that the effective Hamiltonian of Eq.\,(\ref{eqn:hamil_rkky}), derived from the perturbation theory, can also describe the indirect exchange between two spins in the non-perturbative range. 

\begin{figure}[b]
    \includegraphics[width=.49\textwidth]{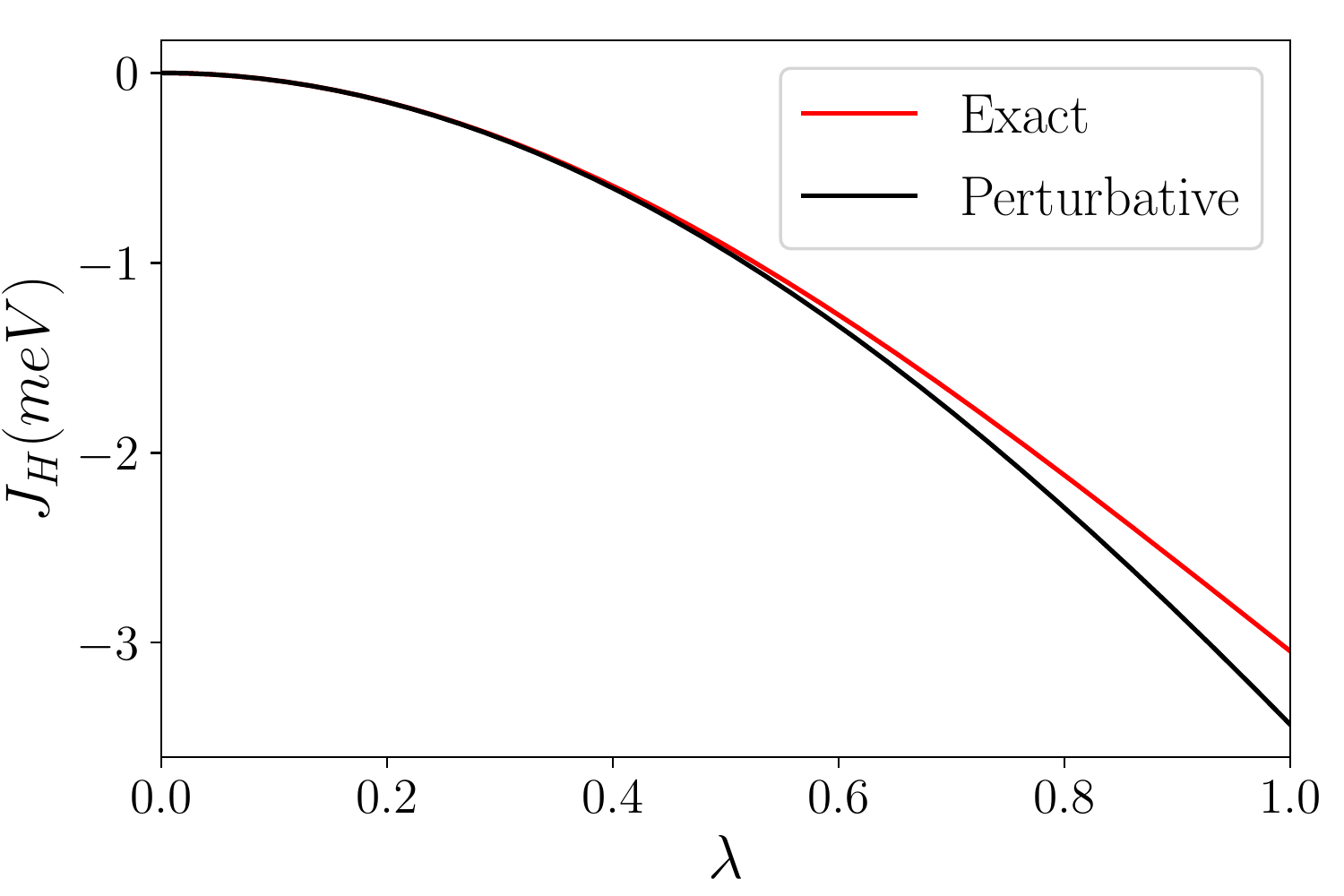}
    \caption{Indirect exchange interaction for two first-neighbor impurities along the zigzag direction as a function of $\lambda=J/t$. The red curve corresponds to the exact calculation computed using Eq.\,\ref{jeff}. The black curve indicates the results obtained from second-order perturbation theory. The parameters of the Hamiltonian [Eq.\,\ref{H0}] are set to $t=1.1$ eV, $\Delta=t_{KM}=0$, and the Fermi energy is equal to $-1$ eV.}
    \label{fig:pert_exact}
\end{figure}

First, we compare the results of perturbation theory with the exact calculation. The figure of merit of the perturbation theory is the ratio $\lambda = J/t$. Hence, we plot the effective coupling obtained from both methods versus $\lambda$ to determine the range of validity of the perturbative results. We take two first-neighbor impurities such that the distance between them is equal to the lattice constant $a$, and, both belong to the same sublattice. For simplicity, we ignore the SOC term as well as the sublattice symmetry-breaking term, $\Delta$, in the Hamiltonian of Eq.\,(\ref{H0}). Thus, the resulting interaction takes the Heisenberg form, $J_H \vec{n}_1\cdot\vec{n}_2$. To compute the exact effective coupling, we calculate the energy difference between the parallel and antiparallel configurations (see [Eq.\,(\ref{jeff})]). In the perturbative case, we calculate the spin susceptibility via the computation of the spin density using Eq.\,(\ref{eqn:spin_density}), and then we obtain the effective interaction by multiplying it by the Kondo coupling squared $J^2$. The resulting curves are plotted in Fig.\,\ref{fig:pert_exact}. As we can see, for $\lambda<0.5$ both methods give similar results. Above this value, the contribution of higher order terms in $\lambda$ becomes significant, so that the exact calculation of the effective coupling becomes necessary.

Now that we have determined the range of validity of the perturbation theory, we discuss how the Kondo coupling can be obtained. An educated guess of the value of $J$ can be determined by comparing experimental measurements of the Yu-Shiba-Rusinov (YSR) states with the ones obtained from the formula that relates them to  $JS\rho$, where $S$ is the spin of the magnetic impurity and $\rho$ is the density of states at the Fermi energy. Using this approach, estimates of $JS=240$ meV have been obtained for magnetic impurities, presumably Fe, in NbSe$_2$\cite{menard15}. On the other hand, the agreement between the energy bands obtained from Eq.(\ref{H0}) and those obtained using Density Functional Theory (DFT) is achieved for $t\simeq 1$ eV. Therefore, in this case, the use of second-order perturbation theory is allowed. 

We can use a second approach to have a rough estimate of $J$: Intra-atomic exchange energies are in the range of 1-2 eV. Hence, if we assume Kondo exchange is inter-atomic, $J$ should be significantly smaller, in line with the estimate from YSR. As discussed above, for those values, perturbation theory works well. However, it could happen that the relevant Kondo exchange is intra-atomic, in which case $J/t$ would not be small, and exact calculations would become mandatory. The non-perturbative regime could also be reached for inter-atomic exchange in spin-valley coupled materials with narrow bands, such as T-TaS$_2$ \cite{yan2015structural}. 

\begin{figure}
     \begin{subfigure}{.48\textwidth}
         \includegraphics[width=248 pt]{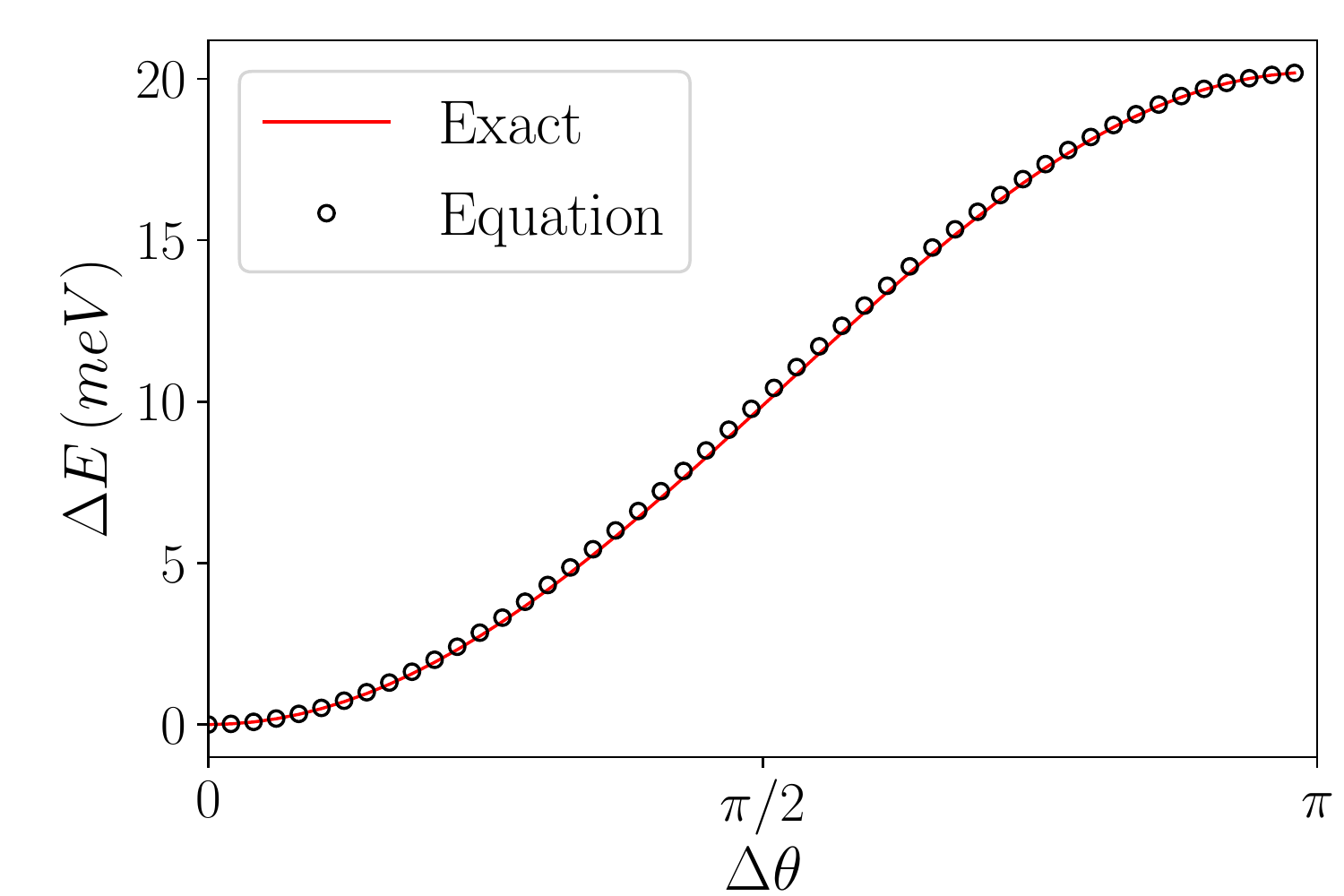}
        \text{\quad \quad \qquad (a)}
         \label{fig:exact_eq_theta}
     \end{subfigure}
     \hfill
     \begin{subfigure}{0.48\textwidth}
         \includegraphics[width=248 pt]{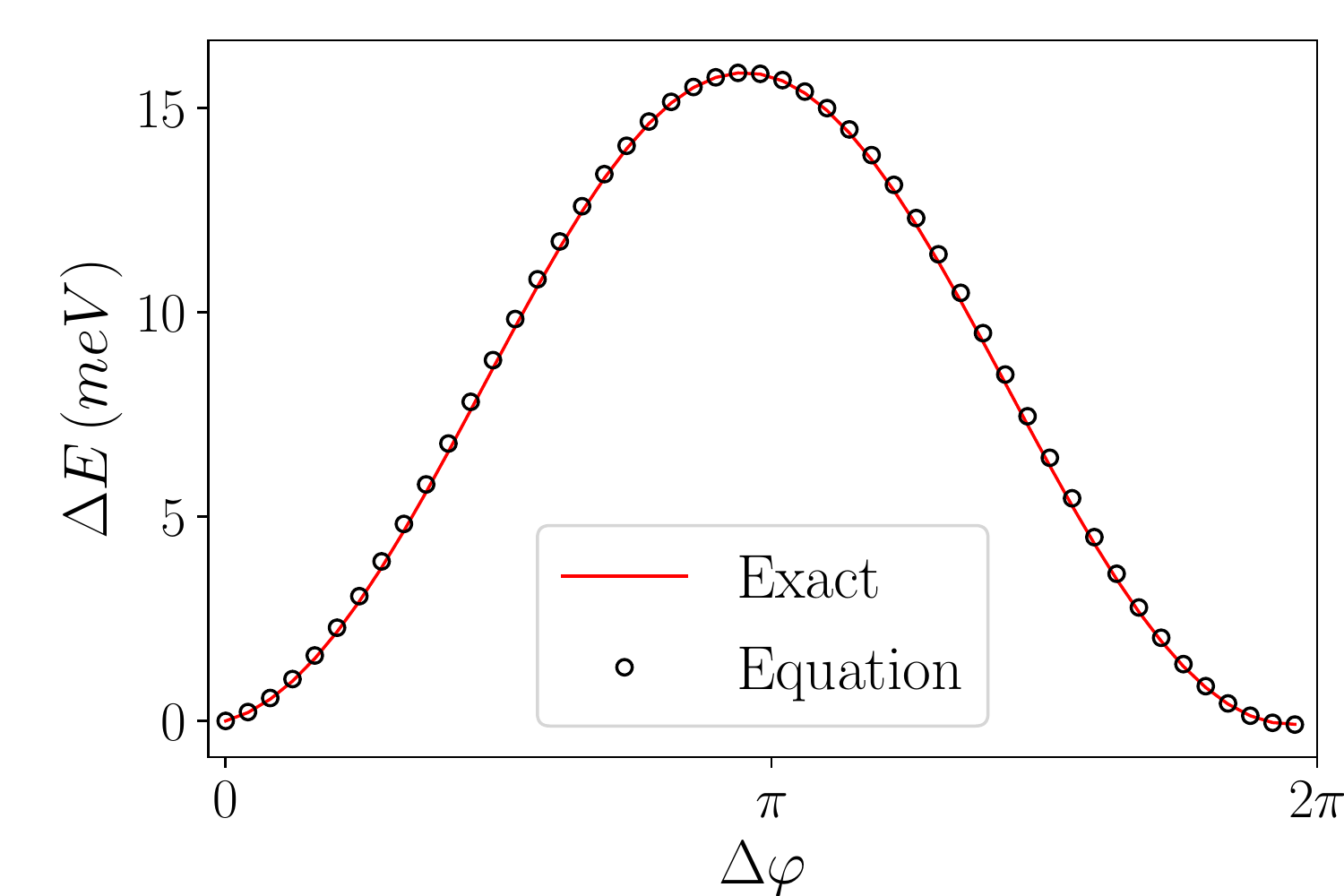}
         \text{\quad \quad \qquad (b)}
         \label{fig:exact_eq_phi}
     \end{subfigure}
\caption{Change of the total energy due to the variation of (a) the polar angle $\theta$, and (b) the azimuthal angle $\varphi$ between the two first-neighbor spins $\vec{n}_1$ and $\vec{n}_2$. Black circles correspond to the results obtained from Eq.\,(\ref{eff_rkky}) and the red line to the ones obtained from the exact diagonalization of the Hamiltonian with two impurities. The Kondo coupling is set to $J=t$, and, we take $t=1.1$ eV, $\Delta=1.66$ eV, $t_{KM}=0.014$ eV in the Hamiltonian [Eq.(\ref{H0})], to  mimic the band structure of MoS\textsubscript{2}.}
\label{fig:exact_eq}
\end{figure}

We have also verified that the effective Hamiltonian [Eq.\,(\ref{eff_rkky})] derived from the perturbation theory works in the non-perturbative range. To do that, we compare results from the equation with those obtained from the diagonalization of the Hamiltonian with two impurities.
The classical impurity spin is parameterized as
$\vec{n}_i =(\sin{\theta_i}\cos{\varphi_i},\sin{\theta_i}\sin{\varphi_i},\cos{\theta_i})$. Then, to illustrate the equivalence between both results, we study the difference of energy caused by the variation of the angles between the spins, $\Delta \varphi$ and $\Delta \theta$. First, we derive the values of $J_{xx}, J_{xy}, J_{zz}$ using Eq.\,(\ref{jeff}) along the high symmetry directions ($\hat{x},\hat{x}), (\hat{x},\hat{y}),(\hat{z},\hat{z})$. Then, using these parameters, we can calculate the total energies for arbitrary directions of the spins. For the exact results, we diagonalize the Hamiltonian with two impurities changing the orientation. We show the results in Fig.\,\ref{fig:exact_eq}, for $J=t$ and $\epsilon_{F}= -1 $ eV. The agreement is very good, with the absolute error below $0.2$ meV and the relative error always below 14 percent. Hence, we have confirmed that the effective Hamiltonian from Eq.\,(\ref{eff_rkky}) is a good approximation even out of the perturbative regime.

\section{RKKY in spin-valley coupled systems}

We now undertake a detailed study of the RKKY coupling for a pair of impurities embedded in a 2D spin-valley coupled crystal. To illustrate the physics of these kinds of systems, we choose the parameters of our model adequate to describe MoS\textsubscript{2}. Specifically, we take $t=1.1$ eV, $\Delta=1.66$ eV, and $t_{KM}=0.014$ eV. These parameters are chosen to obtain bands similar to those given by first-principles band structure calculations \cite{xiao2012coupled}. We consider substitutional impurities in the transition metal sublattice. As a result, in the Kane-Mele model impurities are always in the same sublattice. We study how indirect exchange coupling behaves for different values of the Fermi energy and for different crystallographic orientations.

\begin{figure}

\begin{subfigure}{0.49\textwidth}

\includegraphics[width=250pt]{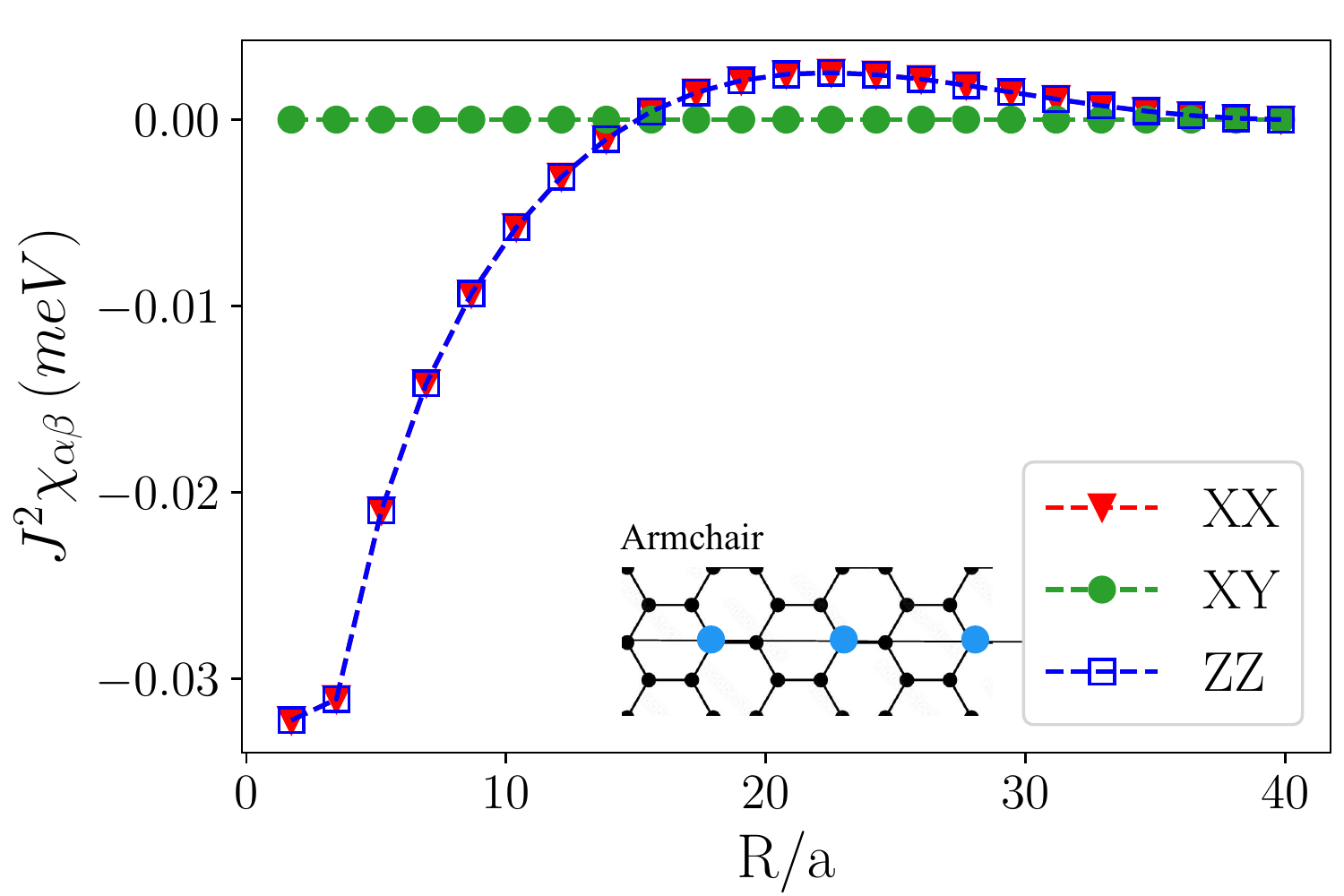}
\text{\quad\quad \qquad (a)}
\label{fig:arm_ef1}
\end{subfigure} 

\begin{subfigure}{0.49\textwidth}

\includegraphics[width=250pt]{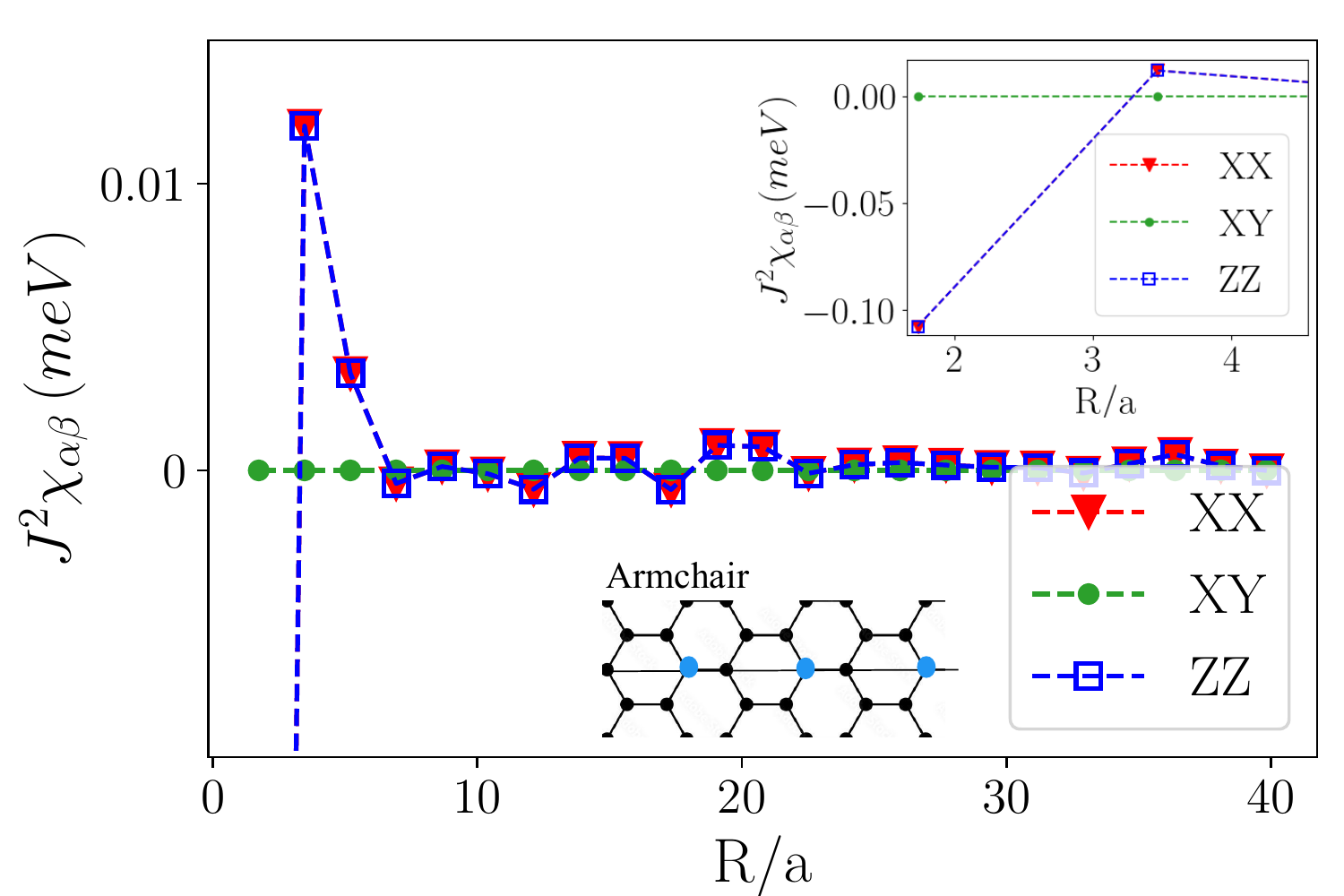}
    \text{\quad\quad \qquad (b)}
    \label{fig:arm_ef2}
    \end{subfigure}
    \caption{Coupling terms of the RKKY interaction,
    $J^2\chi_{\alpha\beta}$, for $J=0.1$t, as a function of impurity separation along the armchair direction.
    Here, $\alpha, \beta = \{XX,XY,ZZ\}$ for the different levels of doping 
    (a)  $\epsilon_{F1} = -0.76$ eV, and, 
    (b) $\epsilon_{F2}= -1$ eV, which are shown in {Fig.\,\ref{fig:system_bands}(b).} The inset displays a zoom of the shorter distances to show the value of the strongest interaction ($\sim 0.1$ meV).}
    \label{fig:direct1}
\end{figure}

\begin{figure}[ht]
    \centering
        
    \begin{subfigure}{0.49\textwidth}
    \includegraphics[width=248pt]{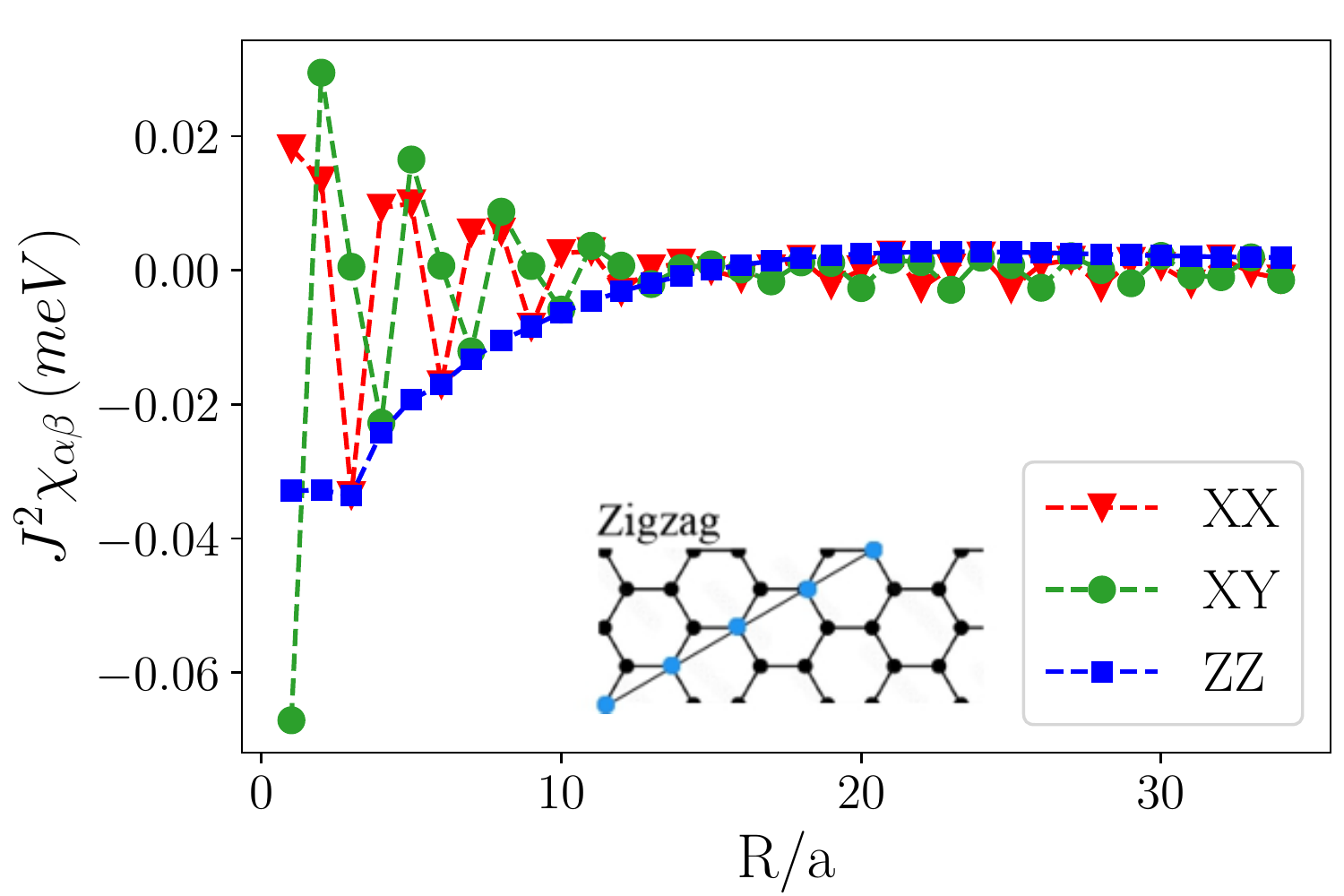}
    \text{\quad \quad  \, \quad \qquad (a) }
    \label{fig:ef1_zig}
    \end{subfigure}
    
    \begin{subfigure}{0.49\textwidth}
    \includegraphics[width=248pt]{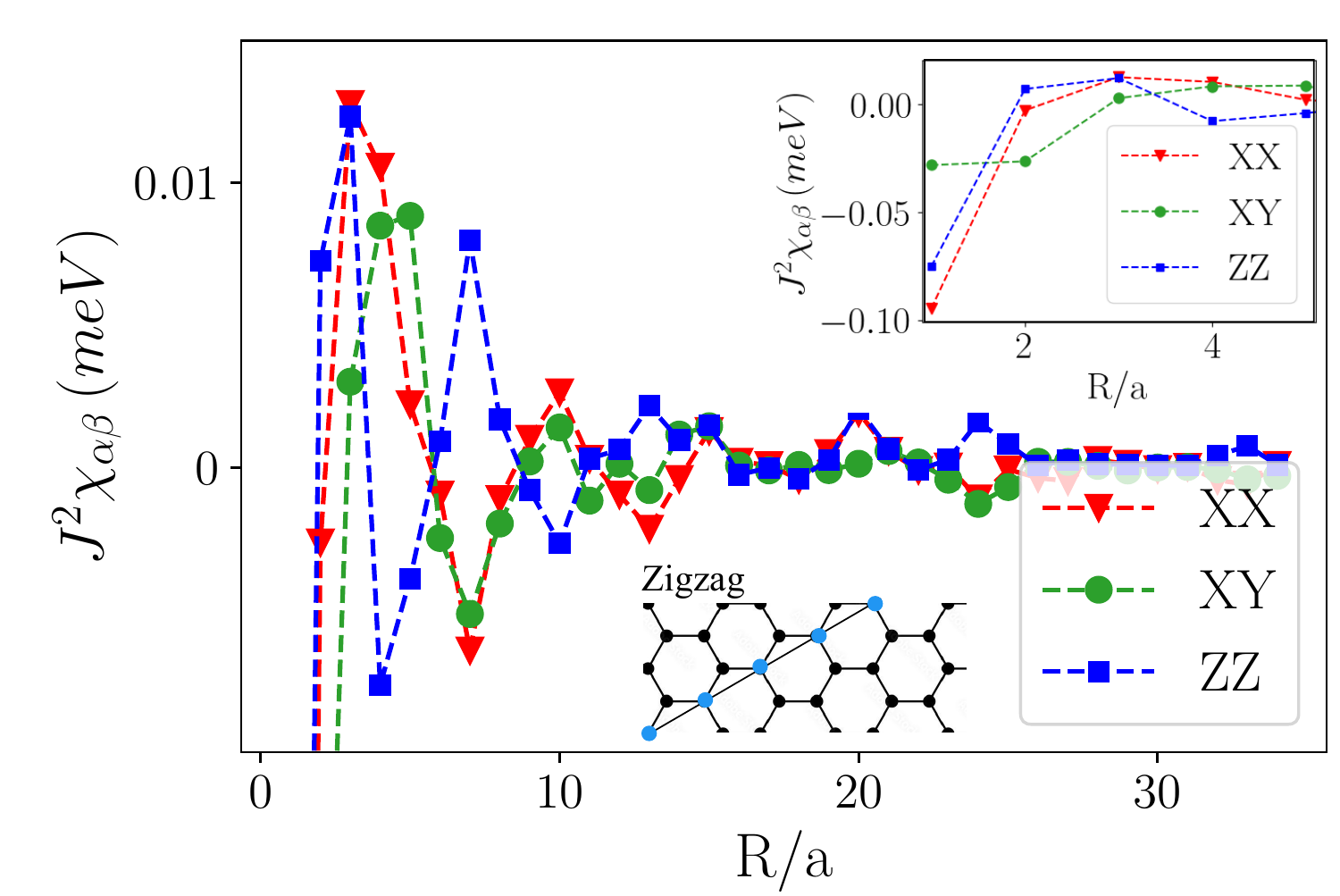}
    \text{\quad \quad  \, \quad \qquad(b)}
    \label{fig:ef2_zig}
    \end{subfigure}
    \caption{Coupling terms of the RKKY interaction, $J^2\chi_{\alpha\beta}$ as a function of impurity separation  along  the zigzag direction, for $J=0.1$t. Here, $\alpha, \beta = \{XX,XY,ZZ\}$ for the different levels of doping (a)  $\epsilon_{F1} = -0.76$ eV and (b) $\epsilon_{F2}= -1$ eV. which are shown in {Fig.\,\ref{fig:system_bands}(b).} The inset displays a zoom of the shorter distances to show the value of the strongest interaction ($\sim 0.1$ meV).}

\label{fig:direct2}
\end{figure}
The isotropic or anisotropic character of the RKKY interaction in spin-valley coupled systems depends on the direction in which impurities are placed. When impurities are placed along the armchair direction, the exchange interaction is isotropic, whereas for the zigzag direction the interaction is anisotropic. In Fig.\,\ref{fig:direct1}, we show the three different couplings of the RKKY interaction for the armchair direction for two different Fermi levels, while in Fig.\,\ref{fig:direct2} we repeat the same plots for the zigzag direction. In the armchair direction, $XX$ and $ZZ$ couplings are equal and the DM vanishes. Thus, the RKYY interaction along the armchair direction is Heisenberg-like, as if the effect of spin-orbit coupling was missing. As we show in the Appendix, the origin of this unexpected spin-rotational symmetry lies in the reflection symmetry of the crystal structure around the armchair direction. It is important to notice that this result does not depend on the Fermi level, and thus it is completely determined solely by the symmetry of the interaction and the lattice of the system. In contrast, when the impurities are placed in the zigzag direction, the three types of the interaction are finite, including the anisotropic Ising and DM terms.

We now study to what extent the behavior of the couplings is affected by the position of the Fermi energy. To do that, we consider two different topologies of the Fermi surface, which are plotted in Fig.\,\ref{fig:system_bands}(c), for each crystallographic orientation. We found that the position of the Fermi energy is related to the magnitude of the indirect exchange and the period of the oscillation of each term of the interaction.  

The maximal value of the interaction for the chosen value of $J=0.1t$ eV is $0.1$ meV for first-neighbor impurities. As we can see in the insets of Figs.\,\ref{fig:direct1}(b) and \ref{fig:direct2}(b), the interaction is stronger for deeper Fermi energy. We attribute this behavior to the increase of the effective density of carriers that can mediate the coupling. We must note though that our single-band prediction may underestimate RKKY compared to multi-band calculations.

We now analyze the period of oscillation of the different terms of the RKKY interaction. We find three different behaviors of oscillation, which we attribute to the scattering processes that contribute to each term of the interaction. The longer period of oscillation is associated with intravalley scattering processes, whereas a shorter period is due to intervalley scattering processes. Thus, we can have oscillations that come from intravalley or intervalley processes but also from a superposition of both of them. We also notice that the intervalley or intravalley nature of the scattering is related to whether the process conserves or flips the spin. 

From the previous considerations, we can explain the behavior of each term of the interaction for the different Fermi levels. Let us first consider the armchair direction. As it can be seen in Fig.\,\ref{fig:direct1}(a), the oscillation period of the indirect exchange is larger for $\epsilon_{F1}$, that has a smaller Fermi wave vector $k_F$, in line with conventional RKKY theory\cite{PhysRev.96.99}. Thus, the processes that contribute to the Heisenberg term are intravalley and conserve the spin. We notice that in the zigzag direction [Fig.\,5(a)] the $ZZ$ component has a similar period to the one obtained for the armchair direction, while the $XX$ component has now a shorter period of $3a$, which corresponds to intervalley contributions. Hence, in this case, the $XX$ term is due to spin-flip processes. We notice that the $XY$ component has the same period but it displays an additional phase shift. On the other hand, for $\epsilon_{F2}$ [Fig.\,5(b)], we observe a complex oscillatory pattern caused by the superposition of oscillations with different wave vectors that come from both intervalley and intravalley scattering processes. 
\begin{figure*}
\centering
\includegraphics[width=\textwidth]{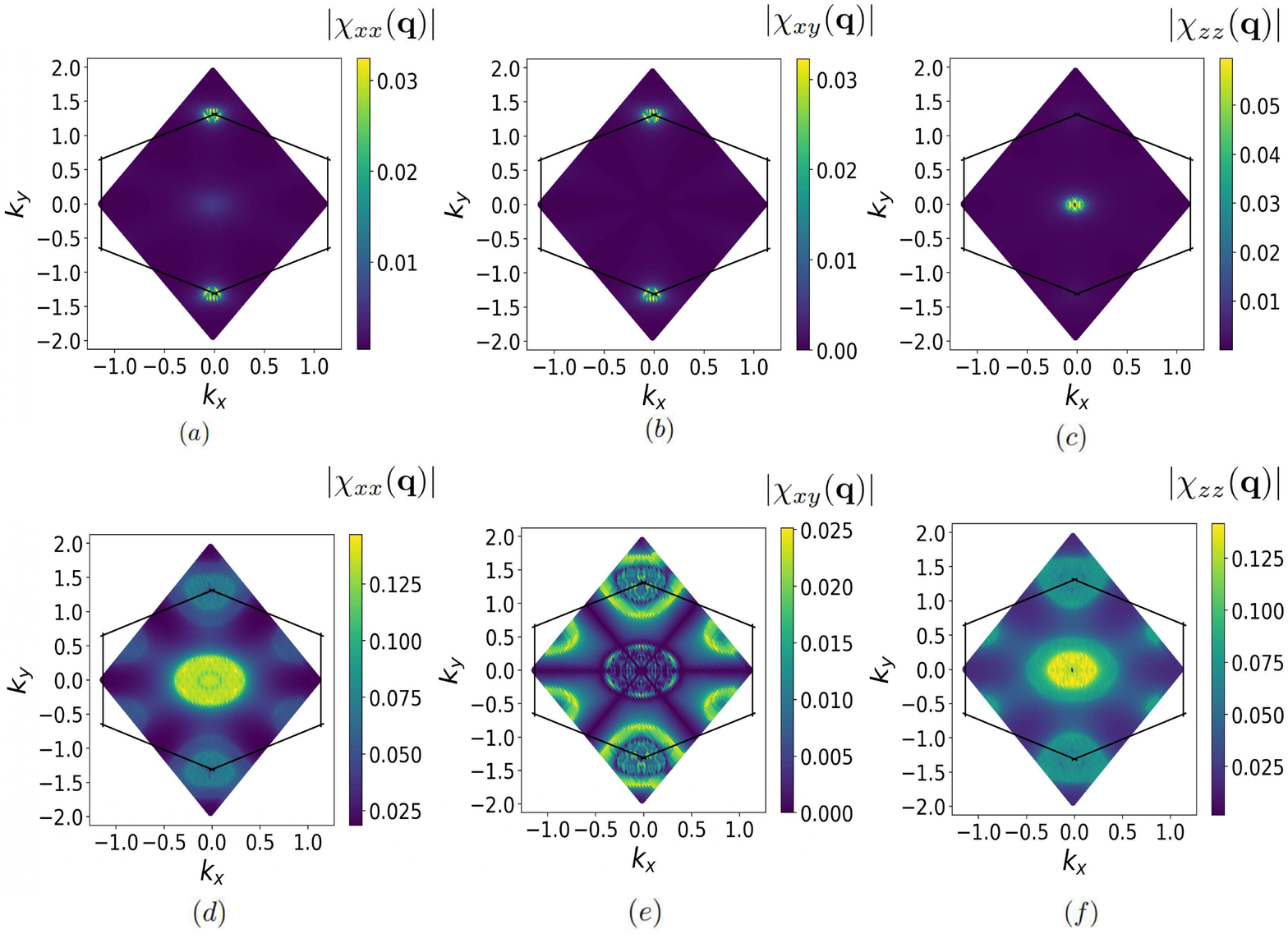}

\caption{Amplitude of the Fourier transform of the elements of the spin susceptibility matrix: $\chi_{xx}$, $\chi_{xy}$, and $\chi_{zz}$: (a)-(c) corresponds to $\epsilon_{F1}= -0.76$ eV; (d)-(f) corresponds to $\epsilon_{F2}= -1$ eV. The solid black line represents the first Brillouin zone. The Dirac points $K$ and $K'$ are located at the corners of the hexagon and the $\Gamma$ point is at the center. The maximum amplitude of the Fourier Transforms located at the $\Gamma$ point represents the intravalley contributions, while the ones located at  $K$ and $K'$ correspond to intervalley scattering processes.}
\label{fig:Fourier_transform}
\end{figure*}
We also analyze the Fourier Transform of the susceptibility matrix elements $\chi(\textbf{q})$. Figures \ref{fig:Fourier_transform} (a)-(c) show the resulting Fourier transforms for the case in which the Fermi level is set to $\epsilon_{F1}=-0.76$ eV. We see that, while the $ZZ$ term only displays intravalley contributions [Fig.\,\ref{fig:Fourier_transform}(b)], $XY$ only presents intervalley spin-flip contributions [Fig.\,\ref{fig:Fourier_transform}(b)]. On the other hand, we notice that the main contribution for the $XX$ term is intervalley. However, it also presents a peak in the $\Gamma$ point, which is caused by the isotropic case in which $ZZ$ and $XX$ terms are equal. In the other case, when we set the Fermi level to $\epsilon_{F2}$, spin-flip and spin-conserving processes can arise from intravalley as well as from intervalley transitions. Thus, the Fourier transforms in this case present both types of contributions, as shown in Figs.\, \ref{fig:Fourier_transform}(d)-(f). However, while intravalley processes highly contribute to $XX$ and $ZZ$ terms, preferred intervalley processes appear in the DM term. Hence, if we can somehow control the system to perform intervalley processes we can thus select DM as the dominant contribution to the indirect exchange.    

\section{Engineering magnetic states: Majumdar-Ghosh Model}
The results of the previous section show that
symmetry, magnitude, and signs of the 
indirect exchange interactions can be very different depending on the Fermi level, the crystallographic orientation, and the distance between impurities. In this section, we study how to leverage this tunability to engineer the Majumdar-Ghosh Model.  

\begin{figure}
\includegraphics[width=0.49\textwidth]{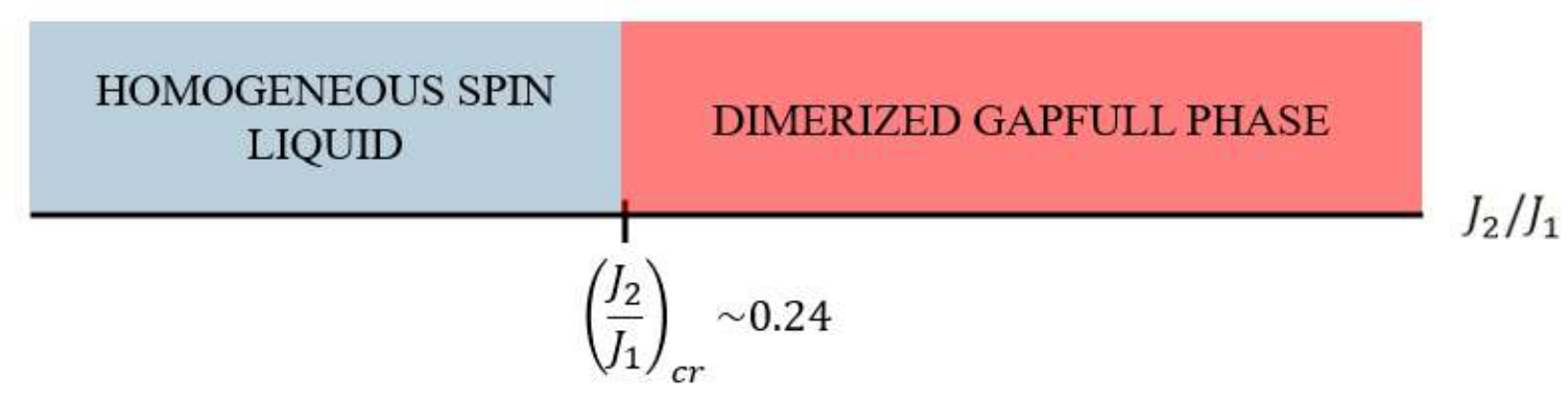}
\caption{Schematic phase diagram of the Majumdar-Ghosh Model. The control parameter is the ratio between $J_2$ and $J_1$. }
\label{fig:phase_diag}
\end{figure}

The Majumdar-Gosh Model is an extended Heisenberg model\cite{majumdar1970antiferromagnetic} that includes both a first and second neighbor antiferromagnetic exchange: 

\begin{equation}
    H =  J_1 \sum_{j = 1}^N \vec{n}_j \cdot \vec{n}_{j+1} + J_2 \sum_{j = 1}^N \vec{n }_j \cdot \vec{n}_{j+2},
\end{equation}
where $J_1$ and $J_2$ stand for nearest-neighbor and next-nearest neighbor. For $J_2=\frac{J_1}{2}$
the ground state of the model can be obtained analytically, and it represents a product state of singlets formed by adjacent spins. As there are two ways to cover the 1D lattice with first-neighbor singlets, the ground state has a twofold degeneracy. The dimerized state can be reached not only at $J_2/J_1 = 1/2$, but in a finite interval of $J_2/J_1$ \cite{PhysRevB.25.4925} from a critical value equal to $(J_2/J_1)_{cr}\approx 0.24$ \cite{okamoto1992fluid}. In the range of interest, it can be distinguished between two phases: the homogeneous spin liquid $J_2/J_1<(J_2/J_1)_{cr}$, and the dimerized gapfull phase at $J_2/J_1>(J_2/J_1)_{cr}$, as shown in the phase diagram of Fig.\,\ref{fig:phase_diag}. 

Now, we explore how to realize the dimerized phase of the Majumdar-Gosh model using magnetic dopants in transition metal dichalcogenides. Since the model has only Heisenberg interactions, our first choice is to consider spin chains along the armchair direction, where anisotropic and DM exchanges vanish. Then, we should look for cases in which $J_1$ and $J_2$ Heisenberg exchange couplings are both antiferromagnetic and the third neighbor coupling is negligible. In order
to achieve these conditions, we play with two degrees of freedom: the lattice constant of the spin chain and the Fermi level. 
In Table.\,\ref{tab:table} we show several instances in which the conditions are fulfilled. In most of them, the ratio $J_2/J_1$ is larger than 0.24, so the dimerized phase would be realized, and in one case the ratio is very close to the value in which the exact analytical solution of Majumdar and Gosh is valid.

\begin{table}[h]

\begin{ruledtabular}
\begin{tabular}{cccccccc}
 Lattice constant [a]&$\epsilon_{F}$\,(eV)&$J_2/J_1$\\
\hline
$\sqrt{3}$ & -1.45 & 0.04\\ 
\hline
$2\sqrt{3}$  & -1.37 & 0.47 \\ 
\hline
$3\sqrt{3}$  & -1.42 & 0.88 \\
\hline
$4\sqrt{3}$  & -1.43 & 0.75\\
& -1.39 & 0.7 \\
\hline
$5\sqrt{3}$  & -1.49 & 0.29\\
& -1.47 & 0.09\\
\end{tabular}
\caption{\label{tab:table}
The ratio between the nearest and next-nearest neighbor couplings, $J_2$ and $J_1$, for the different spin-lattice constants along the armchair direction and Fermi levels for which the Majumdar-Ghosh model is realizable. For values of $J_2/J_1$ above the critical value $0.24$ we get the dimerized ground state, and, for lower values, the ground state corresponds to the spin liquid phase.}
\end{ruledtabular}
\end{table}
\section{Conclusions}
We have studied indirect exchange interactions in spin-valley coupled systems. We go beyond previous work\cite{parhizgar2013indirect,hatami2014spin,mastrogiuseppe2014rkky} in two aspects: we have used a Kane-Mele Hamiltonian without cut-offs in momentum space, and we have treated the exchange between itinerant electrons and local classical moments exactly, by means of exact diagonalization. Our main findings are:
\begin{enumerate}
    \item In the range where perturbation theory is expected to work the indirect exchange is very small, in the range of 0.1 meV, on account of the small value of $J/t$. The fact that indirect exchange obtained in DFT calculations\cite{gao19} is in the range of tens of meV for first-neighbor impurities indicates that either the Kondo exchange in these systems is large so that perturbation theory should be used carefully or not at all, or that electron-electron interactions and/or multiband effects, ignored so far, play an important role enhancing the magnitude of the indirect exchange. This issue will be the subject of future work. 
    \item The difference between perturbation theory and exact is small for values of $J/t<0.5$.
    \item The effective Heisenberg-Ising-DM Hamiltonian of Eq.\,(\ref{eqn:hamil_rkky}), works to a very good approximation in the non-perturbative regime.
    \item The symmetry of indirect exchange is radically different depending on the relative orientation between the crystal and the line that joins the local moments. In particular, interactions along the armchair direction are strictly spin-rotational invariant, with no Ising or DM coupling, due to the reflection symmetry across the armchair direction.
    \item The period of oscillations is determined by the wave vector related to the contributing scattering processes that can be either intervalley or intravalley. These contributions depend on whether the spin of the mediating quasiparticles is conserved or flipped. 
\end{enumerate}
Thus, it is apparent that TMDs provide an extremely versatile platform to engineer indirect exchange interactions of different symmetry and strength, provided that the electronic density and the position of the magnetic impurities can be controlled.

\begin{acknowledgments}
We acknowledge Programa Operativo FEDER /Junta de Andaluc\'ia --- Consejer\'ia de Transformaci\'on Econ\'omica, Industria, Conocimiento y Universidades,
(Grant No. P18-FR-4834.) The Albaicín supercomputer of the University of Granada is also acknowledged for providing computational time and facilities.
B.B. acknowledges financial support from AEI under project PID2021-125604NB-I00. 
J.F.R.  acknowledges financial support from 
 FCT (Grant No. PTDC/FIS-MAC/2045/2021),
 SNF Sinergia (Grant Pimag),
Generalitat Valenciana funding Prometeo2021/017
and MFA/2022/045 and funding from MICIIN-Spain (Grant No. PID2019-109539GB-C41).
\end{acknowledgments}

\appendix

\section{Origin of spin isotropy of indirect exchange along the armchair direction}

\subsection{Reflection symmetry}
We first demonstrate that 
the eigenvalues and eigenvectors of the Bloch Hamiltonian of Eq.\,(\ref{H0}), 
\begin{equation}
    {\cal H}_0(\vec{k})|\phi_\mu(\vec{k}\rangle=
    E_\mu(\vec{k})|\phi_\mu(\vec{k}\rangle,
    \label{eq:A1}
\end{equation}
satisfy 
\begin{eqnarray}
    E_\mu^\sigma(k_x,k_y) = E_\mu^{\bar{\sigma}}(k_x,-k_y) \nonumber\\  \phi^{\sigma}_\mu(k_x,k_y) =\phi^{\bar{\sigma}}_\mu(k_x,-k_y),
    \label{eq:extrasym}
\end{eqnarray}
on account of the reflection symmetry of the honeycomb lattice across the armchair direction (see Fig.\,1(a)). For that matter, we write the Bloch Hamiltonian matrix as:

\begin{equation}
{\cal H}_0(\phi_1,\phi_2)=
\begin{pmatrix}
\frac{\Delta}{2}+\sigma g(\phi_1,\phi_2)& f(\phi_1,\phi_2) \\
f^*(\phi_1,\phi_2) & -\frac{\Delta}{2}-\sigma g(\phi_1,\phi_2)
\end{pmatrix}
\end{equation}
where $\phi_i=\vec{k}\cdot\vec{a}_i$, with $i=1,2$,
$\vec{a}_1=\frac{a}{2}\left(\sqrt{3},+1\right)$,
$\vec{a}_2=\frac{a}{2}\left(\sqrt{3},-1\right)$,
$f(\phi_1,\phi_2)=t\left(1+e^{i\phi_1}+e^{i\phi_2}\right)$
and $
g(\phi_1,\phi_2) = 2t_\mathrm{KM}\left[ \sin\phi_2 - \sin\phi_1 +2\sin(\phi_1-\phi_2)\right]$.
The eigenvalues are given by
\begin{equation}
    E_{\pm}(\phi_1,\phi_2)=\pm \sqrt{\left(\frac{\Delta}{2}+\sigma g(\phi_1,\phi_2)\right)^2+|f(\phi_1,\phi_2)|^2}
\end{equation}
Now, the transformation
\begin{eqnarray}
\left(   \begin{array}{c} k'_x \\k'_y  \end{array}\right)=
\left(   \begin{array}{c} k_x \\ -k_y  \end{array}\right)
\end{eqnarray}
leads to 
\begin{eqnarray}
\left(   \begin{array}{c} \phi'_1 \\\phi'_2  \end{array}\right)=
\left(   \begin{array}{c} \phi_2 \\ \phi_1 \end{array}\right)
\end{eqnarray}
The functions $f$ and $g$ transform as $f'=f$ and $g'=-g$.
Therefore, if upon the reflection and spin reversal, $\sigma'=-\sigma$ the {\em Hamiltonian matrix} remains the same, and as a result, both the eigenvalues and eigenfunctions are also invariant, leading to Eq.\,(\ref{eq:extrasym}).

\subsection{Effective interaction along the armchair direction}
The RKKY interaction tensor has been defined in Eq.\,(\ref{eqn:hamil_rkky}), in terms
of the spin susceptibility in real space. To calculate the RKKY interaction tensor 
along the armchair direction, we first write it in terms of the spin susceptibility in reciprocal
space,
\begin{eqnarray}
    J_{\alpha\beta}(\vec{R}) = \frac{1}{N}\sum_{\vec{k}}e^{i\vec{k}\cdot\vec{R}}
    \chi_{\alpha\beta}(\vec{k}), 
\end{eqnarray}
where both $J_{\alpha\beta}$ and $\chi_{\alpha\beta}$ are matrices in lattice coordinates indices. We now notice that the armchair direction coincides with the $x$ direction in the coordinates system we have adopted. Thus,

\begin{equation}
    J_{\alpha\beta}(x) = \frac{1}{N}\sum_{k_x,k_y}e^{ik_x x}\chi_{\alpha\beta}(k_x,k_y) = 
    \frac{1}{N}\sum_{k_x}e^{ik_x x}
        \overline{\chi_{\alpha\beta}(k_x)}. 
    \label{eq:Jarmchair}
\end{equation}
here $\alpha,\beta$ can take values $x,y,z$ and
\begin{equation}
   \overline{\chi_{\alpha\beta}(k_x)}\equiv     \left[\sum_{k_y}
    \chi_{\alpha\beta}(k_x,k_y)\right]
\end{equation} is the key quantity that determines indirect exchange along the $x$ direction in real space.

We then write  the general formula for the spin susceptibility:
\begin{equation}
\chi_{\alpha\beta}(\vec{k})= \sum_{\sigma_1,\sigma_2,\xi_1,\xi_2} \sigma_{\sigma_1,\sigma_2}^{\alpha} 
\sigma_{\xi_1,\xi_2}^{\beta} 
\chi^{\sigma_1\sigma_2\xi_1\xi_2}(\vec{k},\Omega=0),
\end{equation}
where the matrix elements of $\chi^{\sigma_1\sigma_2\xi_1\xi_2}$ are given by

\begin{widetext}
 \begin{equation}
     \chi^{\sigma_1\sigma_2\xi_1\xi_2}_{ll'}(\vec{k},\Omega)= 
 \frac{1}{N}\sum_{\vec{q}} \sum_{\mu,\nu=1}^{2N} \phi_{l'\xi_2}^\mu(\vec{q})\phi_{l\sigma_1}^{\mu *}(\vec{q})\phi_{l\sigma_2}^\nu(\vec{k}+\vec{q})\phi_{l'\xi_1}^{\nu*}(\vec{k}+\vec{q})\frac{f[E_\mu(\vec{q})]-f[E_\nu(\vec{q}+\vec{k})]}{\Omega+E_\mu(\vec{q})-E_\nu(\vec{q}+\vec{k})+i0^+},
      \label{eq:matrixelementschi}
 \end{equation}
\end{widetext}
with $l,l'$ label the two sites within the honeycomb unit cell. For each wave vector $\vec{q}$ in the Brillouin zone, there are $4$ solutions corresponding to different combinations of the eigenstates of $\hat{S}_z$, $|\up\rangle$ and $|\dn\rangle$ and the atomic orbitals located on the different sites in the unit cell. The functions $\phi^\mu_{l\sigma}(\vec{q})$ are the projections of the eigenvectors of Eq.\,(\ref{eq:A1})
to site $l$ and spin $\sigma$:

\begin{multline}
        |\phi_\mu(\vec{q})\rangle=
    \sum_{l=A,B} \phi^\mu_{l\up}(\vec{q})|\vec{q}\rangle\otimes|\up\rangle + \phi^\mu_{l\dn}(\vec{q})|\vec{q}\rangle\otimes|\dn\rangle,
    \\ \mu=1,...,4.
\end{multline}
In the presence of Kane-Mele effective SOC, the eigenstates of the Hamiltonian are also eigenstates of $\hat{S}^z$, 
\begin{equation}
\phi_{l'\xi_2}^\mu(\vec{q})\phi_{l\sigma_1}^{\mu *}(\vec{q}) =  \phi_{l'\sigma_1}^\mu(\vec{q})\phi_{l\sigma_1}^{\mu *}(\vec{q})\delta_{\xi_2\sigma_1},
\end{equation}
Thus, for a given wave vector, there are two valence bands and two conduction bands, each with a well-defined value of $\hat{S}^z$, such that we can label the bands by $\eta,\eta'=\{c,v\}$, and $\sigma=\{\up,\dn\}$.
We now define the form factor:
\begin{equation}
    \Lambda_{\eta,\eta';l'l}^{\sigma_1,\sigma_2}
    (\vec{q},\vec{k})\equiv
\phi_{l'\sigma_1}^\eta(\vec{q})
\phi_{l\sigma_1}^{\eta *}(\vec{q}) \phi_{l\sigma_2}^{\eta'}(\vec{k}+\vec{q})
\phi_{l'\sigma_2}^{\eta'*}(\vec{k}+\vec{q}). 
\end{equation}
The matrix elements defined in Eq.\,(\ref{eq:matrixelementschi}) become

\begin{widetext}
 \begin{equation}
         \chi^{\sigma_1\sigma_2\sigma_2\sigma_1}_{ll'}(\vec{k},\Omega)=
 \frac{1}{N}\sum_{\vec{q}}
      \sum_{\eta,\eta'=c,v} 
        \Lambda_{\eta,\eta';l'l}^{\sigma_1,\sigma_2}(\vec{q},\vec{k})
      \frac{f[E_{\eta,\sigma_1}(\vec{q})]-f[E_{\eta',\sigma_2}(\vec{q}+\vec{k})]}{\Omega+E_{\eta,\sigma_1}(\vec{q})
      -E_{\eta',\sigma_2}(\vec{q}+\vec{k})+i0^+}.
      \label{eq:matrixelementschi1}
 \end{equation}
\end{widetext}
Having in mind that
\begin{equation}
    \chi_{xx} = \frac{1}{4}\left( \chi^{+-} + \chi^{-+}\right),
\end{equation}
and
\begin{equation}
    \chi_{zz} = \frac{1}{4}\left( \chi^{++} + \chi^{--}\right),
\end{equation}
to understand the behavior of $J^{xx}$ and $J^{zz}$ we need to analyze the following coefficients,

\begin{align}
   &\chi^{+-}_{ll'}(\vec{k}) = \chi^{\up\dn\dn\up}_{ll'}(\vec{k},0)\\
&\chi^{-+}_{ll'}(\vec{k}) = \chi^{\dn\up\up\dn}_{ll'}(\vec{k},0)\\
&\chi^{\sigma\sigma}_{ll'}(\vec{k}) = \chi^{\sigma\sigma\sigma\sigma}_{ll'}(\vec{k},0)
\end{align}

In what follows, we will show that the mirror symmetry of the KM model with respect to the armchair direction implies:

\begin{itemize}
\item $\overline{\chi_{xx}(k_x)} =\overline{\chi_{zz}(k_x)}$
implying that symmetric exchange is isotropic.
\item $\overline{\chi_{xy}(k_x)}=0$, leading to  a vanishing DM exchange. 
\end{itemize}

As a result, indirect exchange along the armchair direction is described by a Heisenberg model, even in the presence of spin-orbit interaction.

As shown in Eq.\,(\ref{eq:Jarmchair}), the Fourier components of the indirect exchange tensor between two spins placed along the armchair direction can be written as sums over the wave vector component perpendicular to that direction. For instance, in the calculation of $J_{xx}$ we find the sum
\begin{widetext}
    \begin{equation}
            \overline{\chi^{+-}_{ll'}(k_x,k_y)} =  
\frac{1}{N}\sum_{k_y}\sum_{q_x,q_y}
      \sum_{\eta,{\eta'}=c,v} \Lambda_{\eta\eta':l'l}^{\up\dn}(k_x,k_y;q_x,q_y)
      \frac{f[E_{\eta\up}(q_x,q_y)]-f[E_{\eta'\dn}(q_x+k_x,q_y+k_y)]}
      {E_{\eta\up}(q_x,q_y)-E_{\eta'\dn}(q_x+k_x,q_y+k_y)+i0^+}.
      \label{eq:avgchipm}
    \end{equation}
We can make the change of variables $q_y\rightarrow -q_y$ and use the symmetry expressed in Eq.\,(\ref{eq:extrasym}) to obtain,
\begin{eqnarray}
    \overline{\chi^{+-}_{ll'}(k_x,k_y)} =  
\frac{1}{N}\sum_{k_y}\sum_{q_x,q_y}
      \sum_{\eta,{\eta'}=c,v} \phi_{l'\dn}^\eta(q_x,q_y)\phi_{l\dn}^{\eta *}(q_x,q_y) 
      \phi_{l\dn}^{\eta'}(q_x+k_x,-q_y+k_y)\phi_{l'\dn}^{{\eta'}*}(q_x+k_x,-q_y+k_y)
      \times\nonumber\\ 
      \frac{f[E_{\eta\dn}(q_x,q_y)]-f[E_{\eta'\dn}(q_x+k_x,-q_y+k_y)]}
      {E_{\eta\dn}(q_x,q_y)-E_{\eta'\dn}(q_x+k_x,-q_y+k_y)+i0^+}.
\end{eqnarray}
A subsequent change of variables in the summation over $k_y$, $k_y-q_y\rightarrow k_y$ leads to,
\begin{eqnarray}
    \overline{\chi^{+-}_{ll'}(k_x,k_y)} =  
\frac{1}{N}\sum_{k_y}\sum_{q_x,q_y}
      \sum_{\eta,{\eta'}=c,v} \phi_{l'\dn}^\eta(q_x,q_y)\phi_{l\dn}^{\eta *}(q_x,q_y) 
      \phi_{l\dn}^{\eta'}(q_x+k_x,k_y)\phi_{l'\dn}^{{\eta'}*}(q_x+k_x,k_y)\times\nonumber\\ 
      \frac{f[E_{\eta\dn}(q_x,q_y)]-f[E_{\eta'\dn}(q_x+k_x,k_y)]}
      {E_{\eta\dn}(q_x,q_y)-E_{\eta'\dn}(q_x+k_x,k_y)+i0^+}.
      \label{eq:transformedchipm}
\end{eqnarray}
The same steps can be applied to $\chi^{-+}$ to give
\begin{eqnarray}
    \overline{\chi^{-+}_{ll'}(k_x,k_y)} =  
\frac{1}{N}\sum_{k_y}\sum_{q_x,q_y}
      \sum_{\eta,{\eta'}=c,v} \phi_{l'\up}^\eta(q_x,q_y)\phi_{l\up}^{\eta *}(q_x,q_y)
      \phi_{l\up}^{\eta'}(q_x+k_x,k_y)\phi_{l'\up}^{{\eta'}*}(q_x+k_x,k_y)\times\nonumber\\ 
      \frac{f[E_{\eta\up}(q_x,q_y)]-f[E_{\eta'\up}(q_x+k_x,k_y)]}
      {E_{\eta\up}(q_x,q_y)-E_{\eta'\up}(q_x+k_x,k_y)+i0^+}.
      \label{eq:transformedchimp}
\end{eqnarray}
Now consider
\begin{eqnarray}
   \overline{\chi^{zz}_{ll'}(k_x,k_y)} =  \frac{1}{N}\sum_{k_y}\sum_{q_x,q_y}
      \sum_\sigma\sum_{\eta,{\eta'}=c,v} \phi_{l'\sigma}^\eta(q_x,q_y)\phi_{l\sigma}^{\eta *}(q_x,q_y) 
       \phi_{l\sigma}^{\eta'}(k_x+q_x,k_y+q_y)\phi_{l'\sigma}^{{\eta'}*}(k_x+q_x,k_y+q_y)\times\nonumber\\ 
      \times\frac{f[E_{\eta,\sigma}(q_x,q_y)]-f[E_{\eta',\sigma}(k_x+q_x,k_y+q_y)]}{E_{\eta,\sigma}(q_x,q_y)-E_{\eta',\sigma}(k_x+q_x,q_y+k_y)+i0^+},
\end{eqnarray}
\end{widetext}
that appears in the calculation of $J_{zz}$ along the armchair direction. If we apply the
change of variables $q_y+k_y\rightarrow k_y$ to the sum over $k_y$ and compare the result to Eq.\,(\ref{eq:transformedchipm}) and
Eq.\,(\ref{eq:transformedchimp}) we see that 
\begin{equation}
    \overline{\chi^{xx}(k_x,k_y)} = \overline{\chi^{zz}(k_x,k_y)},
\end{equation}
thus confirming that $J^{zz}=J^{xx}$ along the armchair direction. 
Following the same line of reasoning, we now show that $\overline{\chi^{xy}_{ll'}(k_x,k_y)}=0$, 
which implies the DM interaction cancels along the armchair direction. We start by noting that 
\begin{equation}
    \chi^{xy} = \frac{i}{4}\left(\chi^{+-} - \chi^{-+}\right).
\end{equation}
Applying the simultaneous transformations $q_y\rightarrow -q_y, k_y\rightarrow -k_y$ to $\overline{\chi^{+-}_{ll'}(k_x,k_y)}$, defined in Eq.\,(\ref{eq:avgchipm}), gives
\begin{widetext}
    \begin{eqnarray}
    \sum_{k_y}\chi^{+-}_{ll'}(k_x,k_y) =  
\frac{1}{N}\sum_{k_y}\sum_{q_x,q_y}
      \sum_{\eta,{\eta'}=c,v} \phi_{l'\dn}^\eta(q_x,q_y)\phi_{l\dn}^{\eta *}(q_x,q_y) 
      \phi_{l\up}^{\eta'}(q_x+k_x,k_y+q_y)\phi_{l'\up}^{{\eta'}*}(q_x+k_x,q_y+k_y)\times\nonumber\\ 
      \frac{f[E_{\eta\dn}(q_x,q_y)]-f[E_{\eta'\up}(q_x+k_x,q_y+k_y)]}
      {E_{\eta\dn}(q_x,q_y)-E_{\eta'\up}(q_x+k_x,q_y+k_y)+i0^+} = \overline{\chi^{-+}_{ll'}(k_x,k_y)}\Longrightarrow 
      \overline{\chi^{xy}(k_x,k_y)=0}.
\end{eqnarray}
\end{widetext}
We finally note that the $J_{\alpha\beta}(\vec{R}=0)$ is also proportional to the unit matrix. Therefore, the exchange-induced single-ion anisotropy vanishes in this system.

\bibliography{referencias_sobraep}	
\bibliographystyle{bib_sobraep}

\end{document}